# Superchiral hot-spots in "real" chiral plasmonic structures.

C. Gilroy[1], D. Koyroytsaltis-McQuire[1], N. Gadegaard[2], A. Karimullah[1] and M. Kadodwala[1*]

[1] School of Chemistry, University of Glasgow, Glasgow, G12 8QQ, UK

[2] School of Engineering, Rankine Building, University of Glasgow, Glasgow G12 8LT, U.K

## Abstract

Light scattering from chiral plasmonic structures can create near fields with an asymmetry greater than the equivalent circularly polarised light, a property sometimes referred to as superchirality. These near fields with enhanced chiral asymmetries can be exploited for ultrasensitive detection of chiral (bio)molecules. In this combined experimental and numerical modelling study, we demonstrate that superchiral hot-spots are created around structural heterogeneities, such has protrusions and indentations, possessed by all real metal structures. These superchiral hot-spots, have chiral asymmetries greater than what would be expected from an idealised perfect structure. Our work indicates that surface morphology could play a role in determining the efficacy of a chiral structure for sensing.

## Introduction

Using the tools of modern nanofabrication, periodic arrays of complex nanostructures of the same design can be routinely manufactured. Although, derived from the same idealised design, the apparently identical individual nanostructures have unique geometric variations, caused by intrinsic surface roughness or structural defects. The presence of these structural irregularities causes highly localised enhancements of EM fields within the overall near field region[1-3]. It has been accepted that when ensembles of nanostructures are considered within an array, the resulting linear optical response (e.g. reflection/transmission) is dependent upon the average of the individual nanostructures' morphologies. Consequently, an array of real structures can be considered a broadened version of that from an array of idealised structures[4]. However, this spatially averaging argument breaks down when one considers Raman scattering and non-linear optical responses, where observed spectroscopic responses are dominated by contributions from localised hot-spots associated with the geometric roughness and defects[5-7]. In addition, for complex structures that consist of multiple elements, surface roughness can alter the level of inductive coupling between elements, and hence modify optical response.

In this study we have investigated how surface roughness influences the level of the chiral asymmetries of near fields created by the optical excitation of chiral plasmonic structures. Near fields generated by light scattering from nanostructures can, in localised regions of space, possess a greater level of chiral asymmetry than comparable circularly polarised light (CPL), a property sometimes referred to as superchirality. Such fields with enhanced chiral symmetry can be exploited for ultrasensitive detection of chiral (bio)molecules[8-13]. Numerical simulations used in previous studies to understand the chiral asymmetries of these field have relied on idealised models of the chiral structure[11, 13, 14]. In this study we have attempted to understand the influence of surface roughness by using a "real" model for the chiral nanostructure, a gammadion, directly derived from atomic force microscopy images, in periodic numerical simulations. The simulations are validated, by comparison with experimental circular dichroism spectra data. Simulated spectra obtained from the real model are in better agreement with the experimental data. Our study reveals that the idealised model underestimates the level of hybridisation, mediated by inductive coupling, between the arms of the gammadion structure. Significantly, surface roughness results in localised regions, close to the walls of the structure, with chiral asymmetries up to 2-3 timesgreater than those obtained from the idealised model. This suggests that surface roughness plays a role in determining the effectiveness of a chiral plasmonic structure for bio-detection applications.

## Results

### AFM Characterisation of structures

The gammadion structure studied has been chosen because it most closely matches those used in previous chiral sensing studies[8, 11]. Representative atomic force microscopy (AFM) images of the structures are shown in **figure 1 (A, B)**. Key parameters have been calculated by taking the mean of 5 measured values from the micrograph, these are shown on a plane view schematic of the structures in **figure 1 (C)**. The gammadion consists of a central cross whose arms are 130 nm wide and 425 nm long and so the structure fits within a 425 × 425 nm square. The arms connected to the centre cross of the structure are thinner at 110 nm. The spacing between the nanostructures is 375 nm.

The surface roughness of a given area is parameterised by the root-mean-square roughness ($R_q$). For the idealised gammadion nanostructures the $R_q$ will be 0 as the surfaces are planar. For the real gammadion, the $R_q$ calculated at its top face is approximately 3.40 nm. Compare this with the glass substrate with a $R_q \sim$ 0.45 nm.

**Real and Ideal models**

The idealised model (Left handed, LH), which simplifies the geometry and morphology of the structures, is shown in **figure 2 (A)**. Most evident simplifications are that the faces and the vertical profiles of the gammadion are made planar: they do not account for surface roughness, sloping or other morphological defects which occur in the fabrication of the nanostructure arrays.

The real model, shown in **Figure (B)** is constructed directly from an AFM micrograph of a single gammadion structure. The micrograph of the entire array, it is cropped so it contains only one structure, selected arbitrarily, and the glass substrate is removed. The file is converted to a format that can be read by the numerical modelling software. This structure includes morphological defects and so it is described here as the 'real' structure.

Error! Reference source not found. **(C)** shows the profile of the 'real' gammadion model at 0, 50 and 100 nm above the glass substrate. The structure shows significant sloping as the height of the structure increases. Tip convolution can overestimate the lateral dimensions of the nanostructures, making protrusions from the surface appear larger and making holes appear smaller[15]. Therefore, it is likely that the micrograph overestimates the extent of the sloping.

**Experimental and simulated CD spectra**

Both experimental and simulated spectra for both gammadion enantiomorphs are displayed in **figure 3**. As expected, both measured and simulated, spectra for LH and RH structures are equal in magnitude but opposite in sign: they are mirrored around the 0 millidegree line of the plot. The experimental data has four pronounced resonances, labelled I-IV in order of increasing wavelength, that have corresponding features in the simulated spectra derived from both models. The positions of these band for measured and simulated spectra are given in **table 1.**

The level of CD observed in the simulated spectra are approximately an order of magnitude larger than those observed experimentally. A reduction by a factor of 2 can be attributed to the 'chessboard' fabrication strategy, which reduces the writing time of the lithography tool by patterning only half of the substrate (see supplementary information). Further reductions must be due to fabrication defects such as missing nanostructures, or missing parts of the structure which are not accounted for in either the ideal or real models.

It is readily apparent that, predictably, the simulations based on the real model provide the best qualitative agreement with the experimental data. However, while both the real and ideal models replicate the position of IV equally well, only the real model provides good qualitative agreement for bands I, II and III. In particular, the idealised model does not replicate the relative intensities of the three resonances, it also under-estimates the wavelength separation between resonances I and II

**Simulated 3-D field plots**

To rationalise the differences between ideal and real models requires an understanding of the origins of the resonances, which can in part be gained from maps of electric field magnitude $|E|$ and chiral asymmetries. The chiral asymmetry of a field can be conveniently parameterised using optical chirality density (*C*) [16-20]

$$C = \frac{1}{2}(\mathbf{D} \cdot \dot{\mathbf{B}} - \mathbf{B} \cdot \dot{\mathbf{D}}), \qquad (1)$$

where $\mathbf{D}$ is the displacement field, $\mathbf{B}$ the magnetic induction and $\dot{\mathbf{D}}$ and $\dot{\mathbf{B}}$ are their respective time-derivatives.

Three dimensional plots of the $|E|$ and *C* have been generated to aid interpretation. Mapping the entire ranges for the $|E|$ and *C* is ineffective for 3-D visualisation, as then only the field at the boundaries of the simulation are visible. By limiting the mapped fields to only those above a certain threshold, allows localised "hot -spot" regions of both $|E|$ and *C* to be easily observed. Analogous 2-D plots are shown in supplementary information.

3-D field plots for $|E|$ and $C$ have been generated for both incident left and right circularly polarised light (LCP and RCP) at the wavelengths of modes I-IV. $C$ values have been normalised against that of RCP at the wavelength of interest. For $|E|$, maps are produced at two thresholds, 10 and 20 V/m, **figures 4** and **5** , while a threshold of 2.5 was set for the $C$ maps, **figure 6.** For $C$ map the highest and lowest values are given to provide guide to the level of the dynamic range. While the largest value is given for the $|E|$, maps.

The $|E|$ field plots for both the real and idealised models display qualitatively similar behaviour. The fields for resonances I – III are all localised on the nanostructure, while for LH (RH) structures under RCP (LCP) illuminations fields are observed to connect neighbouring structures. This observation is consistent with previous work which has shown that resonances I-III are localised modes, associated with inductive coupling between the constituent elements of the gammadion structure, and resonance IV is a lattice (or Bloch) mode arising from the periodicity of the structures[21].

Additionally, for both ideal and real structures the most intense fields are associated predominately with side walls of the nanostructure. For the real structure there are localised hot-spots associated with protrusions on the top rough Au surface, which are not present for flat top surface of the idealised model.

From the $C$ maps it is apparent that regions with the highest level of chiral symmetry ($C$) are located at the bottom of the gammadion structure, in the vicinity of the Au-quartz interface, for both the ideal and real models. The largest values of $C$ for both models are observed for resonance III, which in previous work has been demonstrated to be the most sensitive for the detection of chiral (bio)materials[14]. In contrast to the $|E|$ maps, there are no analogous hot-spots of enhanced optical chirality associated with protrusions of the rough top surface. However, the maximum $C$ values, for a particular enantiomorph and helicity of light, are up to factor of ~ 3 times greater for the real compared to the ideal structure.

**Discussion**

It is unsurprising that the simulations based on the real structure are in closer agreement with the overall measured experimental spectra. However, the ideal and real models do both replicate mode IV, the lattice mode, equally well. This can be attributed to the fact that the lattice mode will be predominately controlled by the periodicity and symmetry of the array. In contrast the localised modes, I – III, which arise through the inductive coupling of the individual rod elements of the gammadion structure would be sensitive to morphological heterogeneity of the real structures. The presence of structural heterogeneity would be expected to influence the properties of the gammadion through two mechanisms. Firstly, the presence of randomly distributed protrusions and indentations on the surfaces of the gammadion act as a symmetry breaking perturbation. In isolation the ideal gammadion structure placed on a substrate belongs to the $C_4$ point group. The presence of random structural heterogeneity effectively reduces the symmetry to $C_1$. Analogous to the procedure used in molecules, a symmetry analysis of the optical modes of the gammadion structure can be carried. The analysis involves a basis set of 8 vectors, mimicking dipoles, associated with the rods that make up the gammadion structure, **Figure 7 (A)**. For the $C_4$ point group this analysis gives four modes 2A, 2B and 2E, with E modes being doubly degenerate, (**Figure 7(B)**). The two E modes represent symmetric and antisymmetric combinations of the dipoles associated with the rods. Only the E modes would be excited by the incident circularly polarised light and the B mode is optically dark. Under the symmetry breaking perturbation of the surface roughness the degeneracy of the E modes is lifted, causing a splitting in to two A modes, this is illustrated in an energy level diagram, **Figure 7(C)**. This symmetry reducing perturbation would be expected to cause a broadening of the CD resonances.

The differences in the positions and line shapes of modes I -II predicted by ideal and real models cannot be solely justified using the symmetry reducing perturbation argument. Specifically, since surface roughness modifies both the intensities and chiral asymmetries of the near fields, creating hot-spots, one would expect that this would alter the level of inductive coupling between the rod elements of the gammadion structure. Intuitively, one would assume that the presence of hot-spots would enhance inductive coupling between the consistent rods. In previous work[21] it has been shown that the wavelength separation between modes I and III (subsequently labelled $S_{I-III}$) scales with increasing coupling between neighbouring rod elements. The value of $S_{I-III}$ derived from the ideal model (78 nm) is smaller than that observed experimentally (92.7 nm), implying that it underestimates the magnitude of the inductive coupling. In contrast the $S_{I-III}$ obtained using the real model (107.9 nm) is larger than the experimental value.

To summarise the current study illustrates the advantages of using realistic structural models for the numerical modelling of the optical properties of metamaterials. Field maps derived from these "realistic" models reveal the presence of localised regions, or hot-spots, of enhanced $|E|$ and $C$ which have a greater magnitude than the maximum values obtained from idealised models. An important implication of structural heterogeneity creating enhanced "superchiral" hot-spot in the vicinity of nanostructure, is that surfaces roughness could enhance the enantiomeric sensing ability of chiral nanostructures.

## Methods

**Numerical modelling**

Numerical simulations of electromagnetic fields were performed using the COMSOL Multiphysics platform. The nanostructure is surrounded by a cuboid representing a unit cell, with the x- and y- dimensions defining the periodicity of the metamaterial, as calculated from AFM images. A schematic showing the simulation geometry is shown in Error! Reference source not found. The z- dimensions of the cell are sufficiently large ($\geq \lambda_{max}/2$) that near-fields generated by the nanostructures do not extend to any integration surfaces above and below, the total height of the cell is 1600 nm. The unit cell is split up into layers of varying thickness. The top 200 nm is a perfectly matched layer (PML) which absorbs all reflections from the nanoparticle. The surface at 200 nm is the excitation port, from where light originates and its polarisation is specified. 100 nm below the excitation port is an integration surface where reflected power is measured. The gammadion is positioned in the centre of the cuboid. 300 nm from the bottom is another integration surface where transmitted power is calculated. 200 nm from the bottom is the radiation exiting port, followed by a 200 nm PML layer. In order to simulate an array of gammadions, Floquet periodic conditions are applied at the x- and y- boundaries.

The material properties of the structure can then be implemented. To replicate the experimental CD work from the previous section, the gammadion nanostructure is placed onto the glass substrate with a refractive index equal to 1.5 then covered with water with refractive index 1.33.

**Gammadion Sample Fabrication**

The gammadia structures were fabricated using an electron beam lithography process. Quartz glass slides were cleaned under ultrasonic agitation in acetone, methanol and isopropyl alcohol (AMI) for 5 minutes each, dried under $N_2$ flow and exposed to $O_2$ plasma for 5 minutes at 100W. A PMMA resist bilayer (*AllResist* 632.12 50K in Anisole and 649.04 200k in Ethyl Lactate) was then spun at 4000 rpm for 1 minute and baked at 180 °C for 5 minutes in between spins. A 10 nm aluminium conducting layer was evaporated on the substrates using a *PLASSYS MEB 550s* evaporator. Patterns were designed on the CAD software *L-Edit* and written by a *Raith* EBPG 5200 electron beam tool operating at 100 kV. The resist was developed in 3:1 MIBK:IPA solution at 23.2°C for 1 minute, rinsed in IPA (5s) and water before drying under $N_2$ flow. A 5nm nichrome adhesive layer was then evaporated below a 100 nm gold layer. The process was completed with a lift-off procedure in acetone at 50°C overnight and then agitated to remove all remaining resist and excess metal.

## Acknowledgement

The authors acknowledge financial support from the Engineering and Physical Sciences Research Council (EP/P00086X/1 and EP/M024423/1) Technical support from the James Watt Nanofabrication Centre (JWNC). DKM was awarded a studentship by the EPSRC. CG's work was supported by the EPSRC CDT in Intelligent Sensing and Measurement, Grant Number EP/L016753/1. MK acknowledges the Leverhulme Trust for the award of a Research Fellowship.

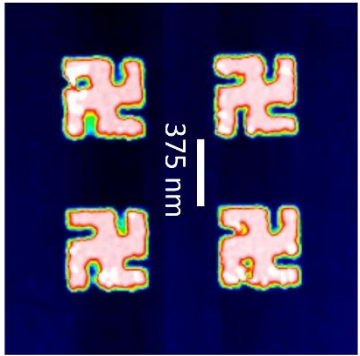

(A)

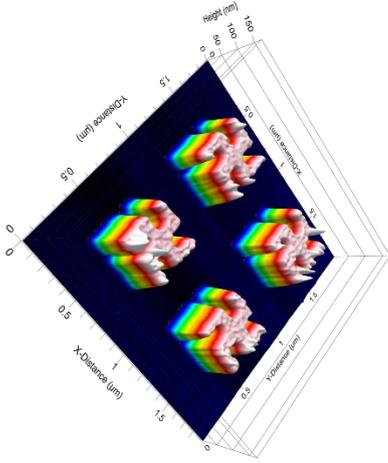

(B)

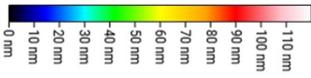

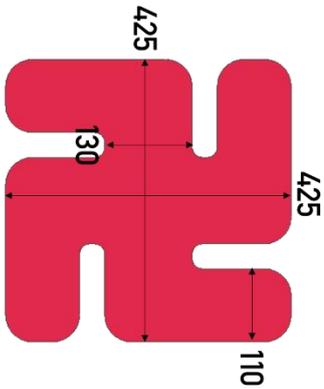

(C)

Figure 1: AFM images of a LH gammadion displayed in (A) plane and (B) 3-D topological views.; (C) is a schematic of the structures which displays relevant dimensions.

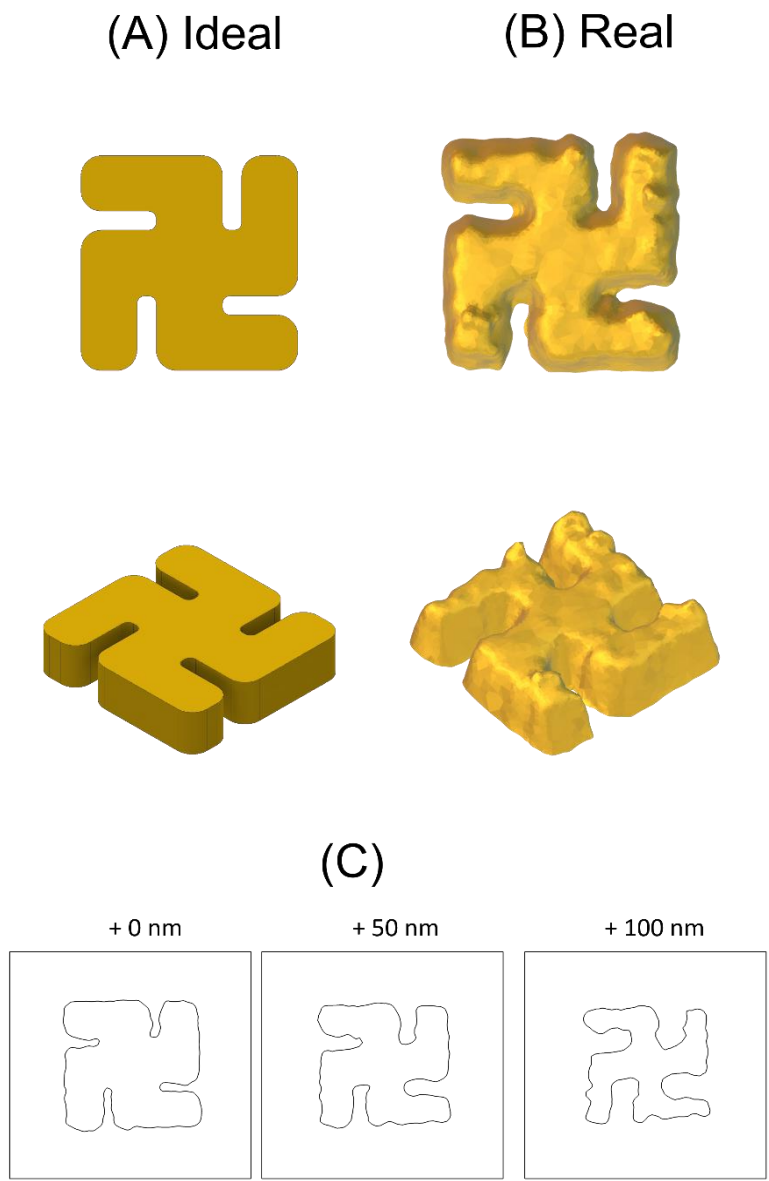

Figure 2: A plan and 3-D view of the (A) ideal and (B) real models of LH gammadion nanostructures used in numerical simulations. (C). Cut slice profile of the 'real' structure 0, 50 and 100 nm above the glass substrate. The structure shows significant sloping.

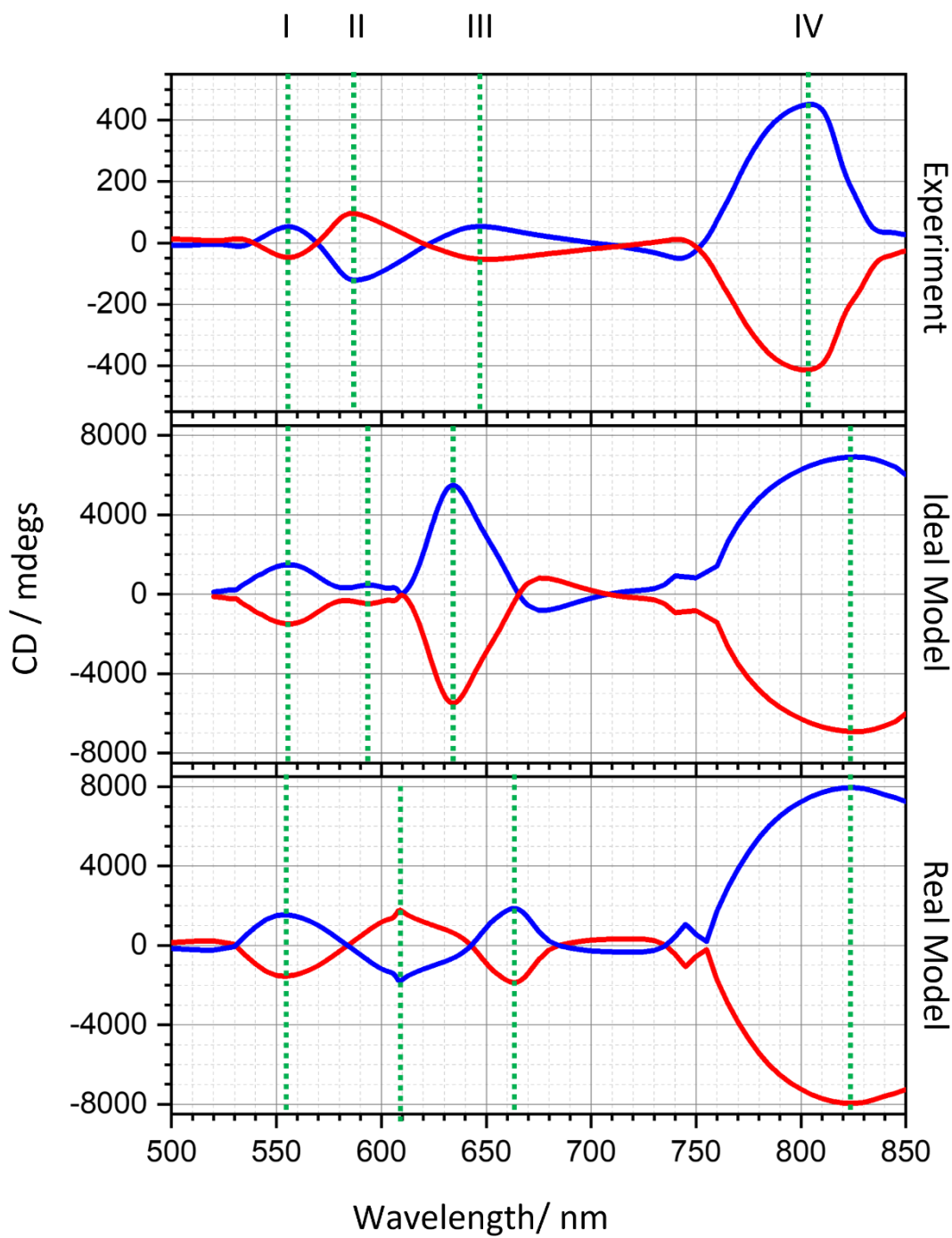

Figure 3: Comparison of the experimental CD spectra (top) for LH (red) and RH (blue) gammadia, with spectra derived from the ideal (middle) and real models (bottom). The positions of modes I-IV are highlighted in the spectra with dotted green lines.

|  | Wavelength Position / nm | | | |
| --- | --- | --- | --- | --- |
|  | I | II | III | IV |
| Experiment (Average LH & RH) | 556.7 | 588.2 | 649.4 | 794.9 |
| Ideal Model | 556.0 | 596.4 | 634.0 | 826.0 |
| Real Model | 555.0 | 608.9 | 662.9 | 824.2 |

Table 1 The (average) experimental positions of modes I-IV are compared with those derived from real and ideal models.

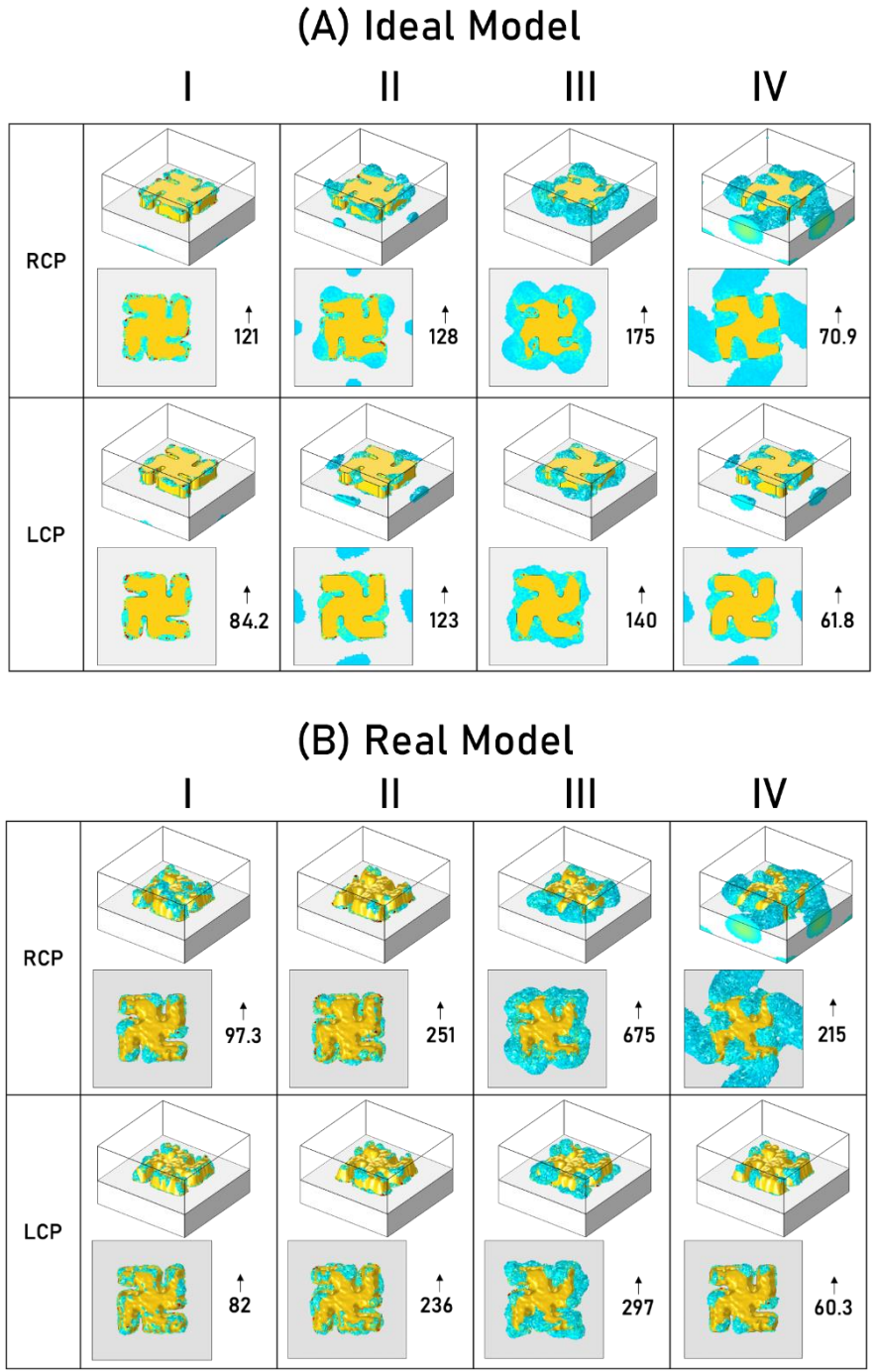

Figure 4: 3-D electric field plots for incident LCP and RCP light, with a 10 V/m threshold, generated from (A) real (B) idealised models. The largest value of $|E|$ is given with an up arrow in each inset.

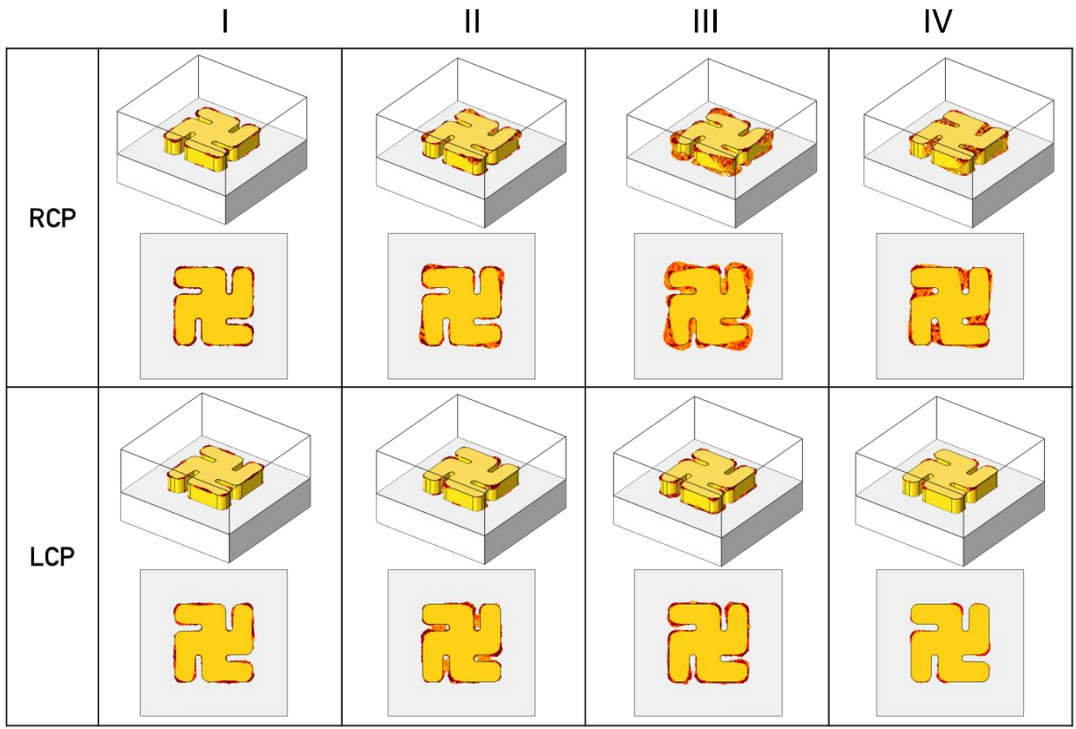

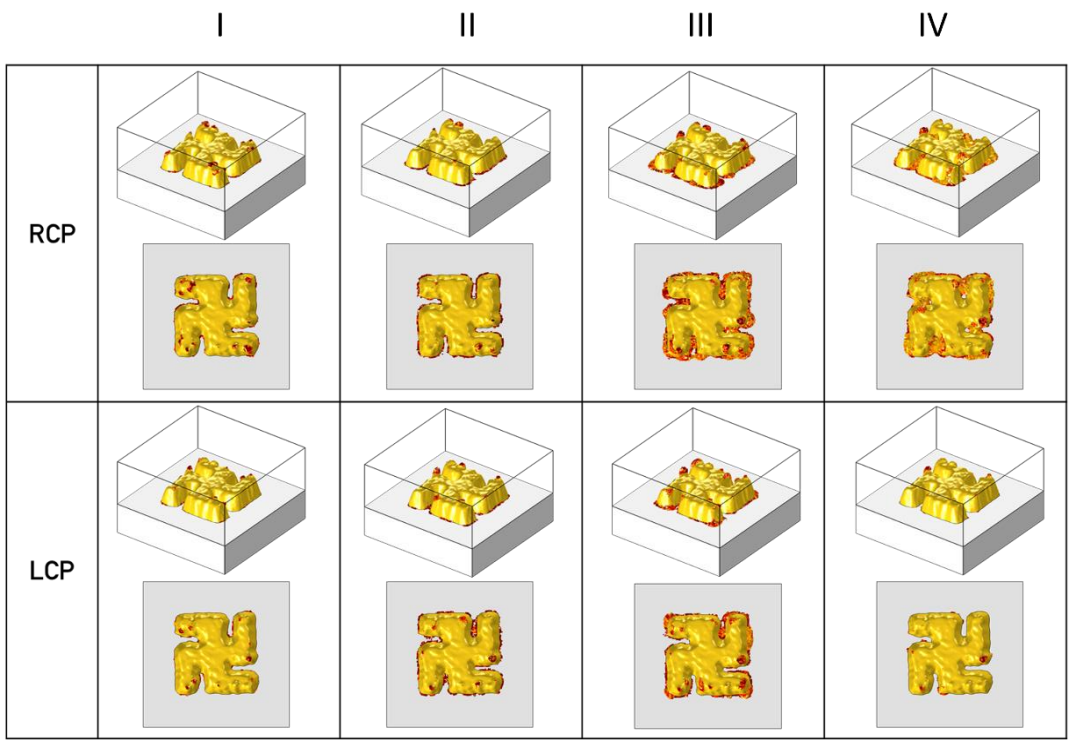

Figure 5: 3-D electric field plots for incident LCP and RCP light, with a 20 V/m threshold, generated from (A) real (B) idealised models

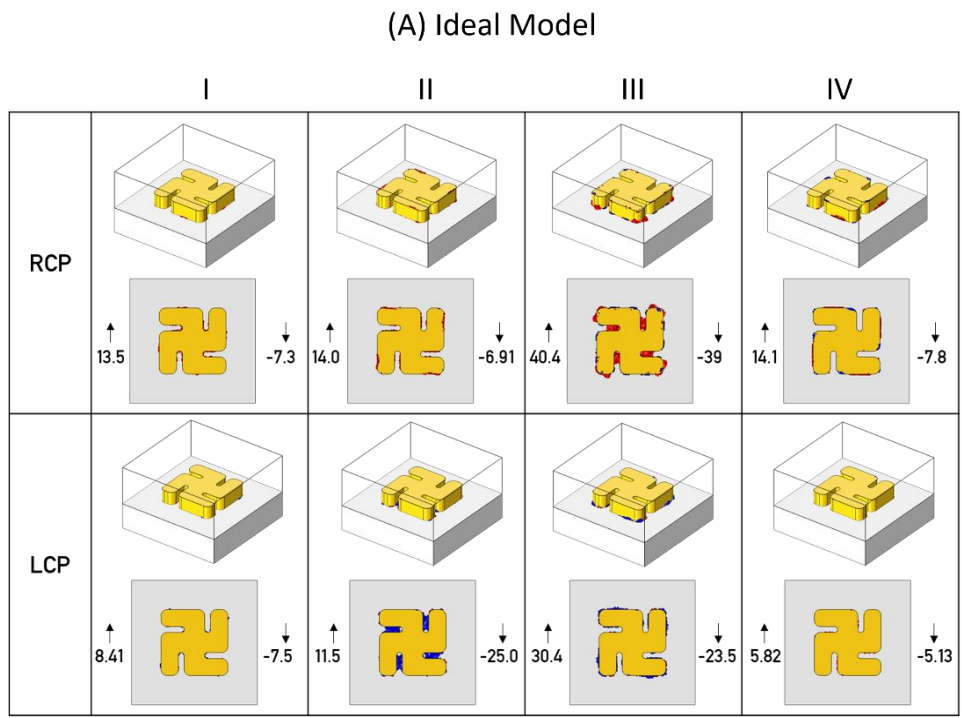

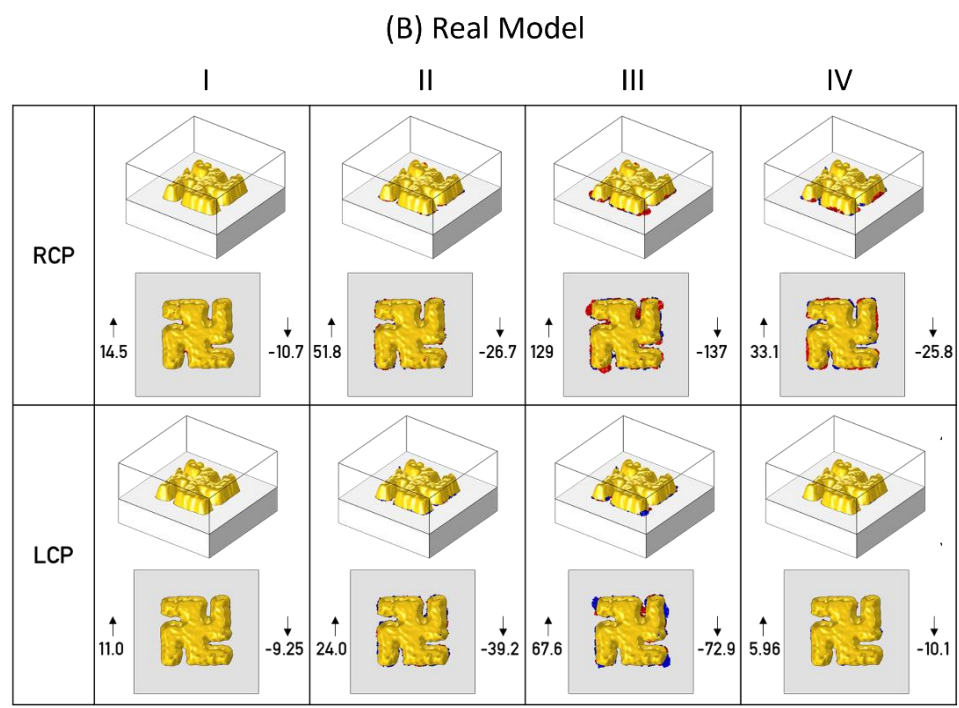

Figure 6: Time averaged normalised (with RCP) optical chirality plots above a threshold of 2.5 (red) (-2.5 (blue)) for (A) real and (B) ideal models. The minimum (↓) and maximum (↑) values for each resonance and light handedness are included.

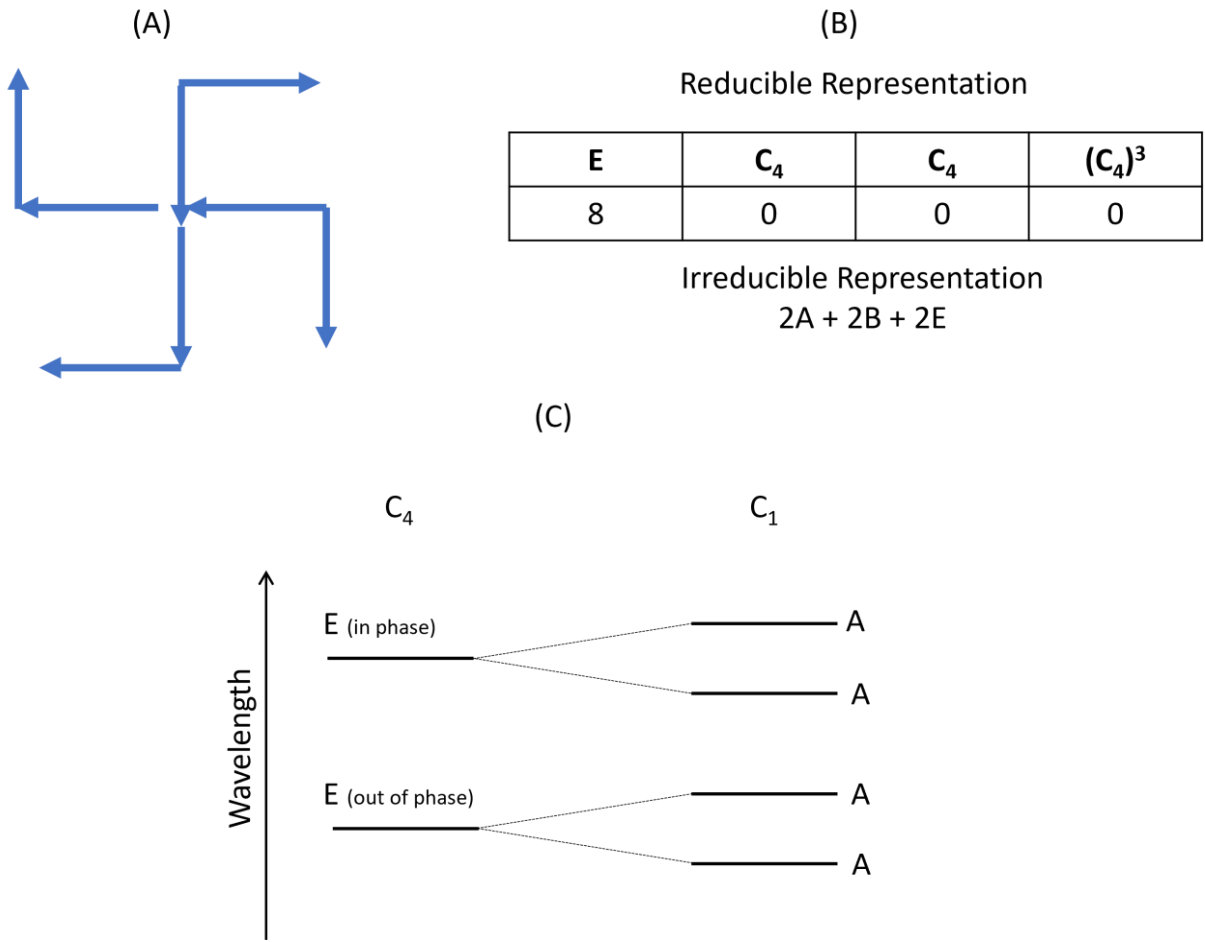

Figure 7: (A) Basis set of vectors used in the symmetry analysis; (B) the reducible and irreducible representations produced by the basis set of vectors. (C) An energy level diagram showing the effects of a symmetry reducing perturbation.

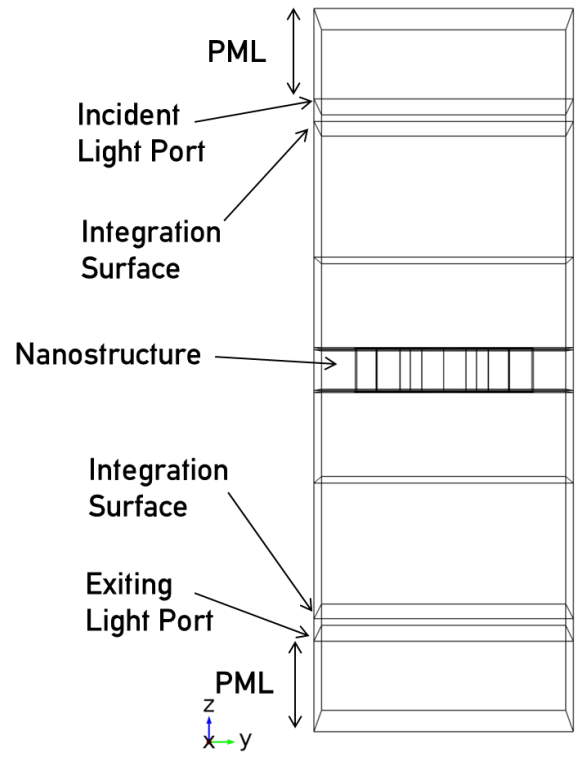

Figure 8: Geometry of the simulation with PML, light ports and integration surfaces identified.